\documentclass[
aps,%
12pt,%
final,%
notitlepage,%
ONESIDE,%
onecolumn,%
nobibnotes,%
nofootinbib,%
superscriptaddress,%
centertags]%
{revtex4}
\usepackage{amsmath}
\usepackage{amssymb}

\begin{document}

\title{NEUTRINO PAIR EMISSION FROM THERMALLY EXCITED NUCLEI IN STELLAR COLLAPSE}

\author{\firstname{Alan~A.}~\surname{Dzhioev}}
\email{dzhioev@theor.jinr.ru} \affiliation{\rm Bogoliubov Laboratory
of Theoretical Physics, JINR, 141980, Dubna, Russia.}

\author{\firstname{A.~I.}~\surname{Vdovin}}
\email{vdovin@theor.jinr.ru} \affiliation{\rm Bogoliubov Laboratory of Theoretical Physics, JINR,
141980, Dubna, Russia.}

\begin{abstract}
We examine the rate of neutrino-antineutrino pair emission by hot
nuclei in collapsing stellar cores. The rates are calculated
assuming that only allowed charge-neutral Gamow-Teller (GT$_0$)
transitions contribute to the decay of thermally excited nuclear
states. To obtain the GT$_0$ transition matrix elements, we employ
the quasiparticle random phase approximation extended to finite
temperatures within the thermo field dynamics formalism. The  decay
rates and the energy emission rates are calculated for the sample
nuclei ${}^{56}$Fe and $^{82}$Ge at temperatures relevant to core
collapse supernovae.
\end{abstract}

\maketitle


\section{Introduction}

The significant role played by processes with neutrinos in
core-collapse supernova is well
known~\cite{Bruenn1991,Fuller1991,Janka2007}. Until the core reaches
the densities $\rho \gtrsim 10^{11}\, \mathrm{g\cdot cm}^{-3}$,
almost all of the energy of the collapse is radiated by neutrinos
that move out of the star freely. Energy emission from the core via
neutrinos helps to maintain the low entropy and, as a result,
nucleons reside primarily in nuclei at the nuclear matter density.
At densities exceeding a few units of $10^{11}\, \mathrm{g\cdot
cm}^{-3}$ high-energy neutrinos become trapped inside the core due
to elastic scattering off nuclei. Due to  this, the lepton fraction
inside the core increases and favors a stronger shock wave
responsible for the supernova explosion. However, low-energy
neutrinos $E_\nu<10$ have the longer mean free path and can
therefore diffuse more easily out of the core. There are a number of
processes contributing to the low-energy neutrino production: the
inelastic scattering of neutrinos on electrons~\cite{Tubbs1975}, the
neutrino-nucleus inelastic scattering~\cite{Bruenn1991}, etc. To
reveal the net effect of neutrinos on the core-collapse process all
relevant neutrino reactions should be included into the supernova
neutrino transport calculations.

In this paper, we examine one of the possible sources of low-energy
neutrinos, namely, the process of the neutrino - antineutrino ($\nu
\bar{\nu}$) pair emission by an excited nucleus
\begin{equation}
(A,Z)^*\to(A,Z)+\nu_k + \bar{\nu}_k.
\end{equation}
Here, the index  $k=e,~\mu,~\tau$ corresponds to three neutrino
flavors.  The astrophysical importance  of this process under the
extreme condition of a stellar collapse (i.e., high temperatures and
matter densities) was first recognized by
Pontecorvo~\cite{Pontecorvo1962}, who noticed that the neutrino-pair
emission by thermally excited nuclei is a powerful mechanism for the
energy loss by stars. It was later pointed out by Bahcall \textit{et
al.}~\cite{Bahcall1974} that despite the low rate this process may
be important because the thermal population of excited nuclear
states can be substantial at stellar temperatures.

Consider  an ensemble of nuclei in equilibrium with a thermal
reservoir  at temperature~$T$.  The role of such a reservoir is
played by the stellar interior and the temperature is in the
neighborhood of 0.5~MeV to 2~MeV ($\mathrm{0.86~MeV}\approx
10^{10}~\mathrm{K}$ = 10\,GK). Nuclear excited states are thermally
populated according to the Boltzmann distribution
$g_i=(2J_i+1)\mathrm{exp}(-E_i/T)$, where $J_i$ and $E_i$ are the
spin and the excitation energy of the level~$i$. We are interested
in the total decay rate from thermally excited nuclear states via
the neutrino-pair emission as well as in the corresponding energy
emission rate. In the nonrelativistic approximation only the
Gamow-Teller (GT) down-transitions contribute to the decay rate
$\lambda_{if}$ from the nuclear state $i$ to the final nuclear state
$f$ with the lower excitation energy $E_f$~\cite{Fuller1991,
Bahcall1974, Crawford1976}
\begin{align}\label{decay_rate}
  \lambda_{if} =&~ 3\frac{G_F^2 g^2_A}{60\pi^3\hbar^7 c^6}\,(\Delta E_{if})^5 B(\mathrm{GT}_0)_{if}
    \notag\\
               =&~3\lambda_0 (\Delta E_{if})^5 B(\mathrm{GT}_0)_{if},~~~\lambda_0\approx 1.72\times 10^{-4}\,\mathrm{s}^{-1}\,\mathrm{MeV}^{-5}.
\end{align}
Here, $\Delta E_{if} = E_i-E_f$ is the transition energy, $G_F$ is
the Fermi weak coupling constant, $g_A\approx1.26$, and
$B(\mathrm{GT}_0)_{if}$ is the reduced transition probability
(strength) associated with the charge-neutral GT$_0$ operator
\begin{equation}
   B(\mathrm{GT}_0)_{if} = (2J_i+1)^{-1}\bigl|\langle i\|\vec \sigma t_0\|f\rangle\bigr|^2,
\end{equation}
where $\vec\sigma$ is the spin operator and $t_0$ is the zero
component  of the isospin operator. The factor of 3 in
Eq.~\eqref{decay_rate} takes into account three neutrino flavors.
Knowing the partial decay rates $\lambda_{if}$, the total decay rate
($\Lambda$) and the energy emission rate ($P$) can be evaluated by
summing over the Boltzmann-weighted, individually determined
contributions
\begin{equation}\label{n_em_rate}
  \Lambda =Z^{-1}(T)\sum_{i,f}g_i\lambda_{if},
\end{equation}
and
\begin{equation}\label{en_loss_rate}
  P=Z^{-1}(T)\sum_{i,f}g_i \Delta E_{if}\lambda_{if},
\end{equation}
where $Z=\sum_i g_i$ is the partition function.

In the pioneering work by Bahcall \textit{et al}.~\cite{Bahcall1974}
the neutrino-pair emission was considered at relatively  low
temperatures when only a few low-lying nuclear states contribute
significantly to the sum $P$~\eqref{en_loss_rate}. The corresponding
Gamow-Teller transition matrix elements between nuclear excited and
ground  states were obtained from the experimental decay rate for
the $M1$ $\gamma$-emission.  At higher temperatures an explicit
state-by-state evaluation of the sums
in~Eqs.~(\ref{n_em_rate})-(\ref{en_loss_rate})  is not feasible
because transitions between nuclear states at high excitations
contribute to the sum significantly\footnote{Using the Fermi gas
model to evaluate the average nuclear excitation energy $\langle
E\rangle \approx AT^2/8$, we find that $\langle E\rangle\sim
30~\mathrm{MeV}$ for the iron-group nuclei ($A\approx50-60$) and for
temperatures $T\sim2~\mathrm{MeV}$.}.
 For such transitions the nuclear level properties
and hence the Gamow-Teller  matrix elements are generally unknown.
In the work by Crawford \textit{et al}. \cite{Crawford1976}, the
neutrino energy emission rate for individual nuclei was
parameterized for temperatures in the range $0.08\le T_{10}\le0.6$
($T_{10}=T/10^{10}~\mathrm{K}$) by replacing the discrete energy
levels with the statistical energy level density when considering
the GT$_0$ transitions from a highly excited state. The GT$_0$
matrix elements were estimated using the method similar to that in
~\cite{Bahcall1974}. Later on, the neutrino-pair emission at high
temperatures was considered by Kolb and Mazurek~\cite{Kolb1979}
making use of the strength function obtained within the Fermi-gas
approach. These calculations were improved by Fuller and
Meyer~\cite{Fuller1991} using the independent single-particle shell
model.

In the present paper, we apply the alternative approach  to study
the neutrino-pair emission from thermally excited nuclei. Our
approach is based on the thermal quasiparticle random phase
approximation (TQRPA). We apply it in the context of thermo field
dynamics (TFD)\cite{TFD1,TFD2,Ojima81} which enables a transparent
treatment of excitation and de-excitation processes and offers the
possibility for systematic improvements. This approach was recently
used in studies of the electron capture and neutrino inelastic
scattering on hot nuclei under supernova conditions~\cite{PRC81,
Dzhioev2011}.

\section{FORMALISM}

The details of our approach  are expounded in~\cite{PRC81,
Dzhioev2011}.  Here, we briefly outline the key points relevant for
the present discussion. In TFD,  a formal doubling of the degrees of
freedom of a nucleus (in the present case) is introduced: a tilde
conjugate operator~$\widetilde A$ (acting in the independent
``tilde'' Hilbert space) is associated with an ordinary operator $A$
(acting in the original Hilbert space) in accordance with properly
formulated tilde conjugation
rules~\cite{TFD1,TFD2,Ojima81}\footnote{The correspondence between
the thermo field dynamics and the superoperator  formalism  used one
of the authors (A.D.) to study nonequilibrium electron transport
phenomena  (see, e.g., \cite{Dzhioev2012}) is discussed
in~\cite{Schmutz1978}.}. The essential ingredients of TFD are the
thermal vacuum $|0(T)\rangle$ and  the thermal Hamiltonian
${\mathcal H}$. The thermal vacuum describes equilibrium properties
of the system. For instance,  the grand canonical average of any
operator can be calculated as the expectation value with respect to
$|0(T)\rangle$. Excited states of the system at finite temperature
correspond to  eigenstates of  the thermal Hamiltonian. The latter
is defined as the difference  between the original system
Hamiltonian $H$ and its tilde counterpart $\widetilde H$, ${\mathcal
H}=H-\widetilde H$. The thermal vacuum is the zero-energy eigenstate
of the thermal Hamiltonian $\mathcal H$ satisfying the thermal state
condition~\cite{TFD1,TFD2,Ojima81}
\begin{equation}\label{TSC}
A|0(T)\rangle = \sigma\,{\rm e}^{{\mathcal H}/2T} {\widetilde
A}^\dag|0(T)\rangle,
\end{equation}
where  $\sigma=1$ for bosonic operators~$A$ and $\sigma=i$
for fermionic operators.

As it follows from the definition of $\mathcal H$, each of its
eigenstates with positive energy has its counterpart -- the
tilde-conjugate eigenstate -- with negative but the same absolute
energy value. Transitions from the thermal vacuum  to positive
energy states (up-transitions) correspond to excitation of the
system, while transitions to negative energy states
(down-transitions) describe decay of thermally excited states. The
transition probabilities  are given by matrix elements of a
transition operator between the thermal vacuum and the excited
state.

In the most practical cases one cannot diagonalize~$\mathcal H$
exactly. Let us assume that we diagonalize the thermal nuclear
Hamiltonian by invoking some approximation methods (e.g., the
mean-field or random phase
approximations~\cite{Hatsuda1989,DzhVdo09}). Then, the thermal
Hamiltonian can be written as
\begin{equation}\label{thHam_approx}
  {\cal H}\approx\sum_i \omega_i(T) ( Q^\dag_i Q_i - \widetilde Q^\dag_i \widetilde Q_i),
\end{equation}
where $\omega_i(T)>0$. The approximate thermal vacuum,
$|0(T)\rangle$,  is the vacuum state for $Q_i,~\widetilde Q_i$
operators. It obeys the thermal state condition~\eqref{TSC}, where
instead  of the exact thermal Hamiltonian we use $\cal H$ given
by~\eqref{thHam_approx}.

Applying the above formalism to the problem of the neutrino-pair
emission  by a hot nucleus we find that the partial decay rate from
the thermal vacuum to the tilde
 state $\widetilde Q_i$ is
given by
\begin{equation}\label{part_DR}
  \lambda_i=3\lambda_0 \omega_i^5\,\widetilde\Phi_i.
\end{equation}
Here, $\omega_i$ is the transition energy (i.e. energy released per decay) and
\begin{equation}\label{amplitude}
 \widetilde\Phi_i=\bigl|\langle\widetilde Q_i\|\vec\sigma t_0\|0(T)\rangle\bigr|^2
\end{equation}
is the GT$_0$ strength for a down-transition. Note, that the  GT$_0$
strength  for  up-transitions, $\Phi_i$, can be obtained
from~\eqref{amplitude} with the help of the detailed balance
principle
\begin{equation}\label{det_bal}
\Phi_i=\bigl|\langle Q_i\|\vec\sigma t_0\|0(T)\rangle\bigr|^2=\exp(\omega_i/T)\widetilde\Phi_i.
\end{equation}
The partial decay rate~\eqref{part_DR} determines the spectrum of
emitted neutrino-antineutrino pairs. Summing over transition energy
gives the total decay rate
\begin{equation}\label{total_DR}
  \Lambda=\sum_i\lambda_i
\end{equation}
and the  energy emission rate
 \begin{equation}\label{energy_ER}
  P=\sum_i\omega_i\lambda_i.
\end{equation}

To obtain the GT$_0$  strength distribution for a hot nucleus we
employ the Hamiltonian of the Quasiparticle-Phonon Model (QPM)
$H_{\rm QPM}$ \cite{sol92}. The QPM Hamiltonian consists of the
proton and neutron mean fields (the part $H_{\rm sp}$), the BCS
pairing interaction $H_{\rm pair}$,  and the separable isoscalar and
isovector particle-hole interaction $H_\mathrm{ph}$
\begin{equation}\label{QPM}
H_{\rm QPM} = H_{\rm sp} + H_{\rm pair}+ H_{\rm ph}.
\end{equation}
The three terms of $H_{\rm QPM}$ read
\begin{align*}
   H_{\rm sp} & = \sum_{\tau=p,n}{\sum_{jm}}^{\tau}(E_{j}-\lambda_\tau)
    a^\dag_{jm}a^{\phantom{\dag}}_{jm}~, \\
H_{\rm pair}& =-\frac14\sum_{\tau=p,n} G_{\tau}{\sum_{\substack{jm \\
j'm'}}}^{\tau}
  a^\dag_{jm}a^\dag_{\overline{\jmath m}}
  a^{\phantom{\dag}}_{\overline{\jmath'm'}}a^{\phantom{\dag}}_{j'm'},
  \\
 H_{\rm ph} &=
-\frac12\sum_{L\lambda}(\kappa^{(L\lambda)}_0+\kappa^{(L\lambda)}_1\vec\tau_1\vec\tau_2)\sum_\mu
S^\dag_{L\lambda\mu} S^{\phantom{\dag}}_{L\lambda\mu},
\end{align*}
where
\[
 S^\dag_{L\lambda\mu} = \sum_{\tau=p,n}{\sum_{\substack{jm \\
  j'm'}}}^\tau
 \langle jm| i^L r^L [Y_{L}\vec\sigma]^\lambda_\mu|j'm'\rangle
  a^\dag_{jm}a^{\phantom{\dag}}_{j'm'}\,.
\]
The quantum numbers $j,m$ actually represent the complete set of
single-particle quantum numbers $n, l, j, m, \tau$ ($\tau = n, p$ is
the isotopic index) and $a_{\overline{\jmath m}}=(-1)^{j-m}a_{j-m}$.
The notation ${\sum}^\tau$  implies the summation over neutron
($\tau=n$) or proton ($\tau=p$) single-particle states only. Since
we assume that only GT$_0$ ( $J^\pi=1^+$)  transitions  contribute
to the neutrino-pair emission, only the spin-multipole part of the
particle-hole interaction is taken into account (actually, we need
only the spin-multipole operators with $\lambda=1,~L=0,2$). The
parameters $\kappa^{(L\lambda)}_0$ and $\kappa^{(L\lambda)}_1$ are
the strength parameters of the isoscalar  and isovector
spin-multipole forces, respectively.

To determine the thermal behavior of a nucleus governed by the
Hamiltonian (\ref{QPM}), we should diagonalize  the thermal
Hamiltonian $\mathcal{H}_{\rm QPM} = H_{\rm QPM} -
\widetilde{H}_{\rm QPM}$ and find the corresponding thermal vacuum
state. Within TQRPA, this  can be done in two steps.

As the  first step, the sum  of the single-particle and pairing
terms $\mathcal{H}_{\rm BCS}=\mathcal{H}_{\rm sp} +\mathcal{H}_{\rm
pair}$ is diagonalized.  To this end, two subsequent unitary
transformations are made. The first is the usual Bogoliubov $u, v$
transformation from the original particle operators
$a^\dag_{jm},~a^{\phantom{\dag}}_{jm}$ to the Bogoliubov
quasiparticle ones $\alpha^\dag_{jm},~\alpha^{\phantom{\dag}}_{jm}$.
The same transformation is applied to the tilde operators
$\widetilde a^\dag_{jm},\ \widetilde a^{\phantom\dag}_{jm}$, thus
producing the tilde quasiparticle operators
$\widetilde\alpha^\dag_{jm},\ \widetilde\alpha^{\phantom\dag}_{jm}$.
The second, unitary thermal Bogoliubov transformation, mixes
non-tilde  and tilde quasiparticles and introduces thermal
quasiparticles
\begin{align}\label{TBt}
       \beta^\dag_{jm}&=x_j\alpha^\dag_{jm}\!-\!i y_j\widetilde\alpha_{jm}\\
       \widetilde\beta^\dag_{jm}&=x_j\widetilde\alpha^\dag_{jm}\!+\!i
       y_j\alpha_{jm}~~~(x^2_j+y^2=1). \nonumber
\end{align}
The coefficients of both transformations are found by diagonalizing
the single-particle part of~${\cal H}_{\rm BCS}$, i.e.
\begin{equation}\label{diag_TBCS}
{\mathcal H}_{\rm BCS}
\simeq\sum_\tau{\sum_{jm}}^\tau\varepsilon_j(T)
(\beta^\dag_{jm}\beta^{\phantom{\dag}}_{jm}-\widetilde\beta^\dag_{jm}\widetilde\beta^{\phantom{\dag}}_{jm}),
\end{equation}
and demanding that the vacuum of thermal quasiparticles,
${|0(T);{\rm qp}\rangle}$, obeys the thermal state
condition~\eqref{TSC} with the Hamiltonian \eqref{diag_TBCS}. As a
result we obtain the well-known equation of  the finite-temperature
BCS. The coefficients $y^2_j$ determine the average number of
thermally excited Bogoliubov quasiparticles and coincide with
thermal occupation factors of the Fermi-Dirac statistics. Because of
thermally excited Bogoliubov quasiparticles, single-particle
transitions which are blocked at zero temperature become thermally
unblocked at finite temperatures. In particular, these are
single-particle down-transitions from high-lying levels to low-lying
ones.

As the second step of the approximate diagonalization of ${\mathcal
H}_{\rm QPM}$, the particle-hole correlations are taken into account
within the thermal QRPA (TQRPA). To this end, we  diagonalize
${\mathcal H}_{\rm QPM}$ in terms of thermal phonon operators
\begin{equation}
{\mathcal H}_{\rm TRPA}=\sum_{\lambda\mu i}\omega_{\lambda i}(T)
   (Q^\dag_{\lambda\mu i}Q^{\phantom{\dag}}_{\lambda\mu i}
   -\widetilde Q^\dag_{\lambda\mu i}\widetilde Q^{\phantom{\dag}}_{\lambda\mu i}),
\end{equation}
where $\omega_{\lambda i}(T)>0$. The thermal phonon operator is
defined as a linear superposition of various pairs (6 variants in
total) of the creation and annihilation ordinary- and tilde thermal
quasiparticle operators
 \begin{multline}\label{phonon}
  Q^\dag_{\lambda \mu i}=\frac12\sum_\tau{\sum_{j_1j_2}}^\tau
 \Bigl\{\psi^{\lambda i}_{j_1j_2}[\beta^\dag_{j_1}\beta^\dag_{j_2}]^\lambda_\mu +
 \widetilde\psi^{\lambda i}_{j_1j_2}[\widetilde\beta^\dag_{\overline{\jmath_1}}
 \widetilde\beta^\dag_{\overline{\jmath_2}}]^\lambda_\mu +
 2i\,\eta^{\lambda i}_{j_1j_2}[\beta^\dag_{j_1}
  \widetilde\beta^\dag_{\overline{\jmath_2}}]^\lambda_\mu\\
+
 \phi^{\lambda i}_{j_1j_2}[\beta_{\overline{\jmath_1}}\beta_{\overline{\jmath_2}}]^\lambda_{\mu} +
 \widetilde\phi^{\lambda i}_{j_1j_2}[\widetilde\beta_{j_1}
 \widetilde\beta_{j_2}]^\lambda_{\mu} +
 2i\,\xi^{\lambda i}_{j_1j_2}[\beta_{\overline{\jmath_1}}
  \widetilde\beta_{j_2}]^\lambda_{\mu}\Bigr\},
\end{multline}
where $[~]^\lambda_\mu$ denotes the coupling of single-particle
angular momenta $j_1, j_2$ to the total angular momentum $\lambda$.
The tilde phonon operator $\widetilde Q^\dag_{\lambda \mu i}$ can be
obtained from~\eqref{phonon} by applying the tilde conjugation
rules. The thermal phonon operators are considered as bosonic ones,
which imposes the normalization constraint on the phonon amplitudes.
To find the energy and the structure of thermal phonons, we apply
the variational principle under the additional constraint that the
vacuum of thermal phonons obeys the thermal state
condition~\eqref{TSC}. As a result, we  obtain the system of linear
equations for the amplitudes $\psi^{\lambda i}_{j_1j_2},\
\widetilde\psi^{\lambda i}_{j_1j_2},\ \eta^{\lambda i}_{j_1j_2}$,
etc. The solvability condition for this system yields the spectrum
of thermal phonons (details can be found in Ref. \cite{DzhVdo09}).
These constitute the equations for the thermal quasiparticle random
phase approximation. The vacuum of thermal phonons,
$|0(T);\mathrm{ph}\rangle$ is the thermal vacuum in  TQRPA.

Once the structure of thermal phonons is determined, one can
evaluate the transition probabilities from the thermal vacuum to
thermal one-phonon states. As was mentioned above, to describe decay
processes, we need to  take into account down-transitions from the
thermal vacuum  to tilde states. Considering the Gamow-Teller
operator $\vec\sigma t_0$ we find the explicit expression for the
transition probability~\eqref{amplitude} in TQRPA (below
$\lambda=1$)
\begin{align}\label{unnar_str1}
  \widetilde\Phi_{i}=&\frac14\Bigl[
 \sum_\tau{\sum_{j_1j_2}}^\tau
 \langle j_1\|\vec\sigma t_0\|j_2\rangle\times
 \notag\\
 &\times
 \bigl\{u^{(-)}_{j_1j_2}(x_{j_1}x_{j_2}\widetilde w^{\lambda i}_{j_1j_2}-y_{j_1}y_{j_2} w^{\lambda i}_{j_1j_2})+v^{(+)}_{j_1j_2}(x_{j_1}y_{j_2}\widetilde s^{\lambda
 i}_{j_1j_2}-y_{j_1}x_{j_2} s^{\lambda
 i}_{j_1j_2}\bigr)\bigr\}\Bigr]^2.
\end{align}
Here, $\langle j_1\|\vec\sigma t_0\|j_2\rangle$ is the reduced
single-particle matrix element of the GT$_0$ operator;
$u^{(-)}_{j_1j_2}=u_{j_1}v_{j_2} - v_{j_1}u_{j_2}$,
 $v^{(+)}_{j_1j_2}=u_{j_1}u_{j_2} + v_{j_1}v_{j_2}$; and
 $w^{\lambda i}_{j_1j_2}=\psi^{\lambda i}_{j_1j_2}-\phi^{\lambda i}_{j_1j_2}$, $s^{\lambda i}_{j_1j_2}=\eta^{\lambda i}_{j_1j_2}-\xi^{\lambda i}_{j_1j_2}$, etc.
 Substituting $\omega_{\lambda i}$  and $\widetilde\Phi_{i}$ into~\eqref{part_DR}
 we find the  partial decay rates from the TQRPA thermal vacuum to the tilde one-phonon states.
 Then, we can find
 the total decay rate Eq.\,\eqref{total_DR} and the energy
 emission rate Eq.~\eqref{energy_ER}.

\section{Calculations for the hot nucleus $^{56}$Fe and  $^{82}$Ge}

Numerical calculations of neutrino-pair emission rates  have been
performed for the  iron isotope $^{56}$Fe and the neutron-rich
germanium isotope $^{82}$Ge. The iron isotope is one of the most
astrophysically relevant nuclei  for the early presupernova
collapse. Neutron-rich nuclei dominate the nuclear composition at a
later stage of collapse. For the considered nuclei the
single-particle wave functions and energies are calculated with the
spherically symmetric Woods-Saxon potential with the parameters
from~\cite{Chepurnov1967}.  For $^{82}$Ge the sequence of
single-particle levels appears to be close to that from
Ref.~\cite{Cooperstein1984}. Within the independent single-particle
shell-model $^{82}$Ge has all the neutron $pf$-shell orbits filled
with the valence neutrons in the $sdg$ shell and the valence protons
in the $pf$-shell.

The constants of the pairing interaction, $G_{p,n}$, are determined
to reproduce experimental pair energies in the BCS approximation.
The pairing does not affect  the obtained results at temperatures
above the critical one, $T_\mathrm{cr}\approx 0.8$~MeV, when pairing
correlations vanish.

As it was pointed out in the previous section,  $J^\pi=1^+$
excitations in spherically symmetric nuclei are generated by the
spin-multipole part of the particle-hole interaction or, more
precisely, by its spin-monopole and spin-quadrupole components. To
determine the isovector, $\kappa^{01}_1,~\kappa^{21}_1$,  and
isoscalar, $\kappa^{01}_0,~\kappa^{21}_0$, coupling parameters, we
use the standard estimates for the separable schematic forces with
radial form factors $r^\lambda$~\cite{Castel1976}.

\begin{figure}[h]
\includegraphics[width=0.7\textwidth]{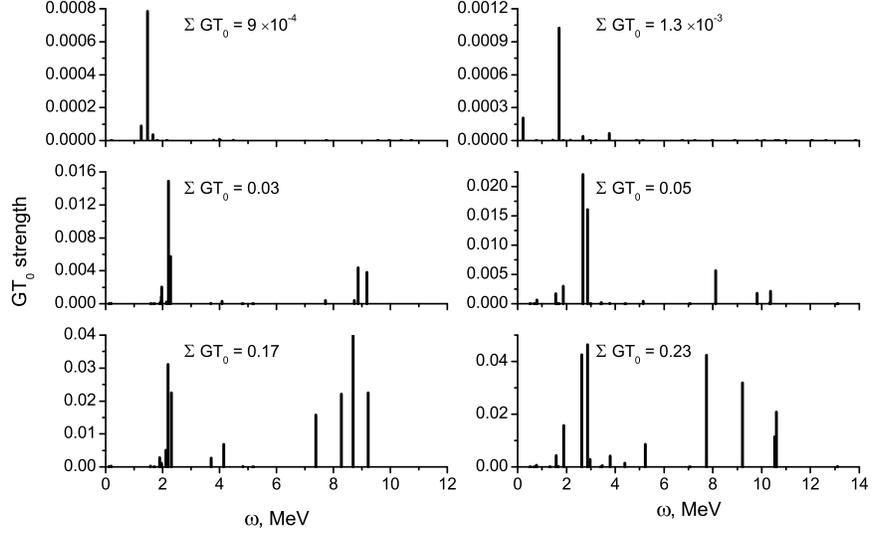}
\caption{The Gamow-Teller strength distribution for down-transitions
in $^{56}$Fe (left panels) and $^{82}$Ge (right panels) at three
values of temperature: $T=0.5$~MeV (upper panels),  $T=1.5$~MeV
(middle panels), and  $T=2.5$~MeV (lower panels). For each value of
$T$ the total strength of down-transitions is
displayed.}\label{figure1}
\end{figure}

First, we have performed TQRPA calculations of the GT$_0$ strength
distribution in $^{56}$Fe and $^{82}$Ge.   In Fig.~\ref{figure1}, we
display at three different temperatures the strength distributions
of GT$_0$  transitions from the thermal vacuum to tilde one-phonon
states. We recall that only these down-transitions contribute to the
decay process. All figures are plotted as a function of the
transition energy $\omega_i>0$. While the details of the strength
distributions vary between two nuclei, the essential features are
the same. Namely, at low temperatures ($T=0.5$~MeV) the transition
strength is concentrated in a narrow resonance structure within the
energy region between~2~and~3~MeV. With increasing temperature one
observes the increase in the total  GT$_0$ strength accompanied by
the appearance of higher energy transitions. Referring to
Fig.~\ref{figure1}, one sees that the high-energy GT$_0$ transition
strength is more fragmented than the low-energy one.

All these features can be readily explained.  First of all, we note
that  only transitions between the levels which are the spin-orbit
partners contribute to the  GT$_0$ strength. As a consequence, the
transition energies are determined by the spin-orbit splitting in
the nucleus under study. Some increase in the transition energy
comes from the pairing effects and the repulsive particle-hole
interaction.

At low temperatures, single-particle down-transitions between levels
close  to proton and neutron Fermi surfaces  are effectively
thermally unblocked. For $^{56}$Fe these transitions are
$2p_{1/2}\to2p_{3/2}$ for both protons and neutrons. In the
neutron-rich nucleus $^{82}$Ge the neutron level $2p_{3/2}$ is well
below the Fermi level and, therefore, it is mostly filled at low
temperatures. As a result, only the proton single-particle
down-transition $2p_{1/2}\to2p_{3/2}$ is thermally unblocked in
$^{82}$Ge at low $T$. Thus, at low temperatures the energy of
thermally unblocked  GT$_0$ transitions is given by the spin-orbit
splitting of  the  $2p$ shell, which is in the range $2-3$~MeV.

Increasing temperature promotes nucleons to high-lying
single-particle levels.  This makes possible down-transitions
between spin-orbit partners with a large spin-orbit splitting. In
$^{56}$Fe these transitions are $1f_{7/2}\to1f_{5/2}$ and
$1g_{9/2}\to1g_{7/2}$ for both neutrons and protons. The
fragmentation of the high-energy GT$_0$ strength, which is most
pronounced at high
 temperatures $T=2.5$~MeV, reflects the distinction in the spin-orbit
 splitting
of $1f$ and $1g$ shells. The same is true for $^{82}$Ge, with the
only difference  that the $1h_{11/2}\to1h_{9/2}$ neutron transition
contributes to the GT$_0$ strength above 10~MeV.

\begin{figure}[h]
\includegraphics[width=0.7\textwidth]{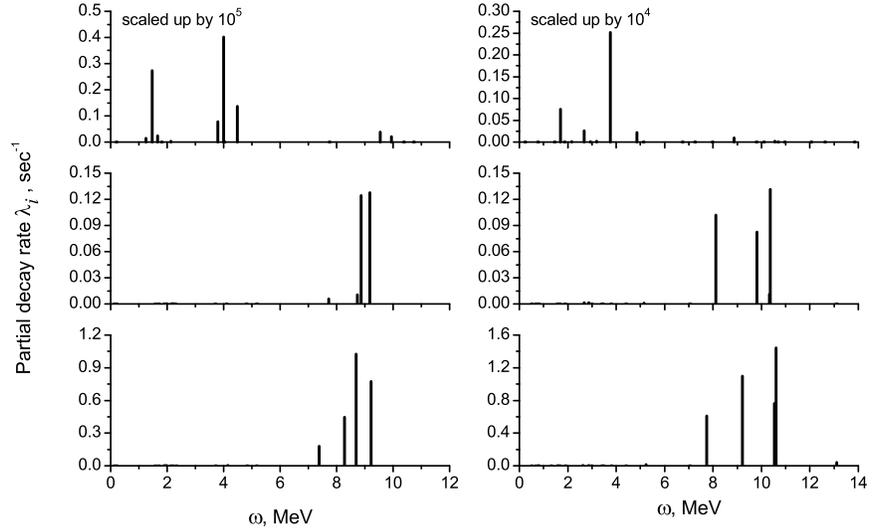}
\caption{Partial decay rates~\eqref{part_DR} for $^{56}$Fe  (left
panels) and $^{82}$Ge (right panels) as functions of the transition
energy. The partial decay rates are shown for three values of
temperature: $T=0.5$~MeV (upper panels),  $T=1.5$~MeV (middle
panels) and  $T=2.5$~MeV (lower panels). }\label{figure2}
\end{figure}

According to Eq.~\eqref{part_DR}, the transition strengths should be
multiplied  by the fifth power of the transition energy to give
partial decay rates or, what is the same, the spectrum of emitted
neutrino pairs. Figure~\ref{figure2} shows partial decay rates
computed in this fashion for considered nuclei at temperatures
$T=0.5,~1.5$~and~$2.5$~MeV. As is obvious from the figure,
multiplying by $\omega_i^5$ clearly favors high-energy transitions.
The effect is most pronounced at intermediate temperatures.
Comparing Figs.~\ref{figure1} and~\ref{figure2},  we see that at
$T=1.5$~MeV the GT$_0$  strength  for down-transitions is mostly
concentrated  in the range $2-3$~MeV, whereas the spectrum of the
emitted neutrino-pair peaks at much higher energies. Thus, we
conclude that the temperature growth increases the energy of emitted
neutrinos.  We will return to this point at the end of this Section.

\begin{figure}[h]
\setcaptionmargin{30mm} \onelinecaptionsfalse
\includegraphics[width=0.5\textwidth]{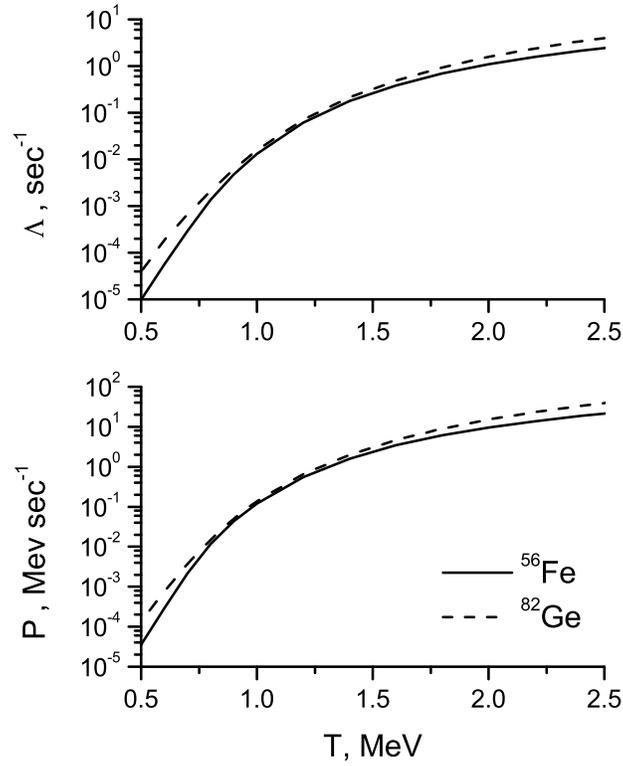}
\caption{The total decay rate $\Lambda$ (upper panel) and the energy
emission rate $P$ (lower panel) for $^{56}$Fe  and $^{82}$Ge as
functions of $T$.}\label{figure3}
\end{figure}

In Fig.~\ref{figure3}, the total decay rates
$\Lambda$~\eqref{total_DR} and energy emission rates
$P$~\eqref{energy_ER} for $^{56}$Fe and $^{82}$Ge are shown as
functions of $T$. As expected,  both the rates demonstrate a strong
temperature dependence. From the above discussion it becomes clear
that the main reason for this is the thermal unblocking of
high-energy GT$_0$ down-transitions, since  more strength at higher
$\omega$ means faster emission rates. Moreover, we find that at the
considered $T$ values the calculated rates in $^{56}$Fe and
$^{82}$Ge are close to each other despite a considerable difference
in their masses. This result is not surprising since in both the
nuclei the same single-particle down-transitions dominate the
neutrino-emission process. Comparing the obtained energy emission
rate for $^{56}$Fe with earlier estimates~\cite{Fuller1991,Kolb1979}
we find that our results are very close to those obtained
in~\cite{Fuller1991} by applying the independent single-particle
shell-model.

\begin{figure}[h]
\onelinecaptionstrue
\includegraphics[width=0.5\textwidth]{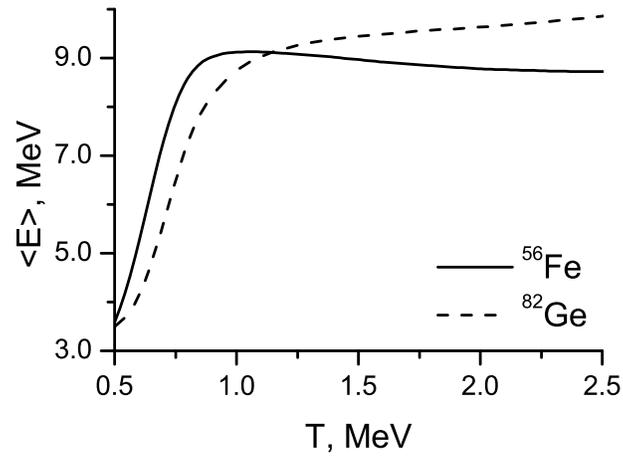}
\caption{The mean energy $\langle E \rangle$ of $\nu \bar\nu$ pairs
emitted from $^{56}$Fe and $^{82}$Ge as a function of temperature
$T$.}\label{figure4}
\end{figure}

Given the energy emission rates and  the total decay rates we can
find a mean  energy of emitted neutrino pairs as $\langle E\rangle =
P/\Lambda$. The results of these calculations are shown in
Fig.~\ref{figure4}. As evident from the figure, at low $T$ the mean
energy of the emitted neutrino-pairs is less than 9~MeV and $\langle
E\rangle$ rises rapidly with temperature till $T\approx 1.0$ MeV.
After that $\langle E\rangle$ becomes nearly independent of $T$.
Thus, for $T> 1.0$ MeV the mean energy of the emitted neutrino-pairs
is determined by the spin-orbit splitting of $f,~g$ and $h$ shells.
The last observation is important for neutron-rich nuclei with
$A\sim100$. Here, we would like to stress that for $T>2.0$~MeV the
contribution of the fist-forbidden high-energy transitions is not
negligible and should be taken into account (see
Ref.~\cite{Fuller1991}). By doing so we expect to observe an
increase in $\langle E\rangle$ for $T>2.0$~MeV.

\section{Conclusions}

We have performed studies of a neutrino-antineutrino pair emission
process  from thermally excited nuclei. The studies are relevant for
core collapse supernova simulations, as the considered process may
play an important role in production of low-energy neutrinos and,
hence, in the energy transport during the collapse. As an example,
the neutrino-pair emission from  $^{56}$Fe and $^{82}$Ge was
considered for temperatures $0.5\le T\le 2.5$~MeV by taking into
account the allowed GT$_0$ transitions only.  Thermal effects were
treated within the thermal quasiparticle random phase approximation
in the context of the thermo field dynamics formalism.

Throughout the paper, we discussed the underlying nuclear physics of
this process in terms of the single-particle down-transitions
between the spin-orbit partner levels. It was found that the $f,~g$
and $h$ shell spin-orbit transitions dominate the processes for
$T\gtrsim 1$MeV, whereas the $p$ shell spin-orbit transitions are
important only at lower $T$. Our calculations revealed the same
effect as was found in~\cite{Fuller1991,Kolb1979}: A temperature
increase leads to a considerable enhance of the neutrino-pair
emission rate.  According to our calculations, this enhancement is
due to the thermally unblocked $f,~g$ and $h$ spin-orbit
down-transitions. The last observation is important for neutron-rich
nuclei.

The calculated energy emission rate for $^{56}$Fe is close to that
obtained in~\cite{Fuller1991} within the independent single-particle
shell-model. To extend our calculations to higher temperatures, we
need to take into account the first-forbidden transitions. This will
be the subject of our future studies.

\newpage

\centerline{Figure captions}
\begin{itemize}
\item[\bf Fig.1]
The Gamow-Teller strength distribution for down-transitions
in $^{56}$Fe (left panels) and $^{82}$Ge (right panels) at three
values of temperature: $T=0.5$~MeV (upper panels),  $T=1.5$~MeV
(middle panels), and  $T=2.5$~MeV (lower panels). For each value of
$T$ the total strength of down-transitions is displayed.

\item[\bf Fig.2]
Partial decay rates~\eqref{part_DR} for $^{56}$Fe  (left panels) and
$^{82}$Ge (right panels) as functions of the transition energy. The
partial decay rates are shown for three values of temperature:
$T=0.5$~MeV (upper panels),  $T=1.5$~MeV (middle panels) and
$T=2.5$~MeV (lower panels).

\item[\bf Fig.3]
The total decay rate $\Lambda$ (upper panel) and the energy emission
rate $P$ (lower panel) for $^{56}$Fe  and $^{82}$Ge as functions of
$T$.

\item[\bf Fig.4]
The mean energy $\langle E \rangle$ of $\nu \bar\nu$ pairs emitted
from $^{56}$Fe and $^{82}$Ge as a function of temperature $T$.

\end{itemize}

\newpage

\newpage

\end{document}